\documentclass[12pt,preprint]{aastex}

\usepackage{graphicx}
\newcommand{\msun}{\ensuremath{\mathit{M}_{\odot}}}               
\newcommand{\lsun}{\ensuremath{\mathit{L}_{\odot}}}                  

\def\ec{$\eta$~Car}
\def\eca{$\eta_{\mathrm{A}}$}
\def\ecb{$\eta_{\mathrm{B}}$}

\shorttitle{A Lighthouse Effect in $\eta$ Car}
\shortauthors{Madura \& Groh}
\slugcomment{ \today}
\begin{document}

\title{A Lighthouse Effect in Eta Carinae\altaffilmark{1}}

\author{Thomas I. Madura and Jose H. Groh}
\affil{Max-Planck-Institut f\"{u}r Radioastronomie, Auf dem H\"{u}gel 69, D-53121 Bonn, Germany}

\altaffiltext{1}{Based on \emph{HST} ACS/HRC observations.}
\begin{abstract}
We present a new model for the behavior of scattered time-dependent, asymmetric near-UV emission from the nearby ejecta of \ec. Using a 3-D hydrodynamical simulation of \ec's binary colliding winds, we show that the 3-D binary orientation derived by \citet{madura12} is capable of explaining the asymmetric near-UV variability observed in the \emph{Hubble Space Telescope} Advanced Camera for Surveys/High Resolution Camera (\emph{HST} ACS/HRC) F220W images of \protect\citet{smith04b}. Models assuming a binary orientation with $i \approx 130^{\circ}$ to $145^{\circ}$, $\omega \approx 230^{\circ}$ to $315^{\circ}$, $\mathrm{PA}_{z} \approx 302^{\circ}$ to $327^{\circ}$ are consistent with the observed F220W near-UV images. We find that the hot binary companion does not significantly contribute to the near-UV excess observed in the F220W images. Rather, we suggest that a bore-hole effect and the reduction of Fe II optical depths inside the wind-wind collision cavity carved in the extended photosphere of the primary star lead to the time-dependent directional illumination of circum-binary material as the companion moves about in its highly elliptical orbit.

\end{abstract}



\defcitealias{madura12}{M12}
\defcitealias{smith04b}{S04}
\defcitealias{hillier01}{H01}
\defcitealias{hillier06}{H06}

\section{Introduction} \label{intro}

Multi-wavelength observations obtained over the past twenty years strongly indicate that $\eta$~Carinae is a highly eccentric ($e \sim 0.9$) colliding wind binary with a $2022.7 \pm 1.3 \ \mathrm{day}$ orbital period \citep{damineli96, pittard02, corcoran05, damineli08a, damineli08b}. With a total luminosity $\gtrsim 5 \times 10^{6}$~\lsun, dominated by the primary star \eca, a luminous blue variable \citep{davidson97}, \ec's total binary mass is $\gtrsim 120$~\msun \citep[][hereafter H01, H06]{hillier01, hillier06}.

One topic that remains the subject of debate is the orientation of \ec's orbit. Most favor an orbit in which the less-massive, hotter companion star \ecb\ is behind \eca\ at periastron \citep{damineli97, pittard02, corcoran05, iping05, hamaguchi07, nielsen07, damineli08b, okazaki08, gull09, gull11, parkin09, groh10b}. However, others place \ecb\ on the near side of \eca\ at periastron \citep[e.g.][and references therein]{falceta09, kashi09}. Settling this debate is crucial as a precise set of orbital parameters is key for determining the individual stellar masses.

\citet[][hereafter S04]{smith04b} attempted to constrain the geometry of \ec's orbit using asymmetric variability seen in near-ultraviolet (NUV) images of the Homunculus nebula obtained with the \emph{Hubble Space Telescope} Advanced Camera for Surveys/High Resolution Camera (\emph{HST} ACS/HRC). Alternating patterns of bright spots and `shadows' observed on opposite sides of \ec\ before and after its 2003.5 spectroscopic event are interpreted by \citetalias{smith04b} as being due to a time-variable, asymmetric NUV radiation field that arises from (1) intrinsic NUV emission from \eca's outer wind and (2) UV radiation from \ecb\ that preferentially escapes in directions away from \eca. In the scenario proposed by \citetalias{smith04b}, a dark shadow appears on the opposite side of \eca\ near periastron because its dense wind blocks \ecb's far-UV radiation over a large, time-varying solid angle. Using this interpretation, \citetalias{smith04b} suggest that \ec's orbital major axis is nearly perpendicular to the observer's line-of-sight, with \ecb\ on the far side of \eca\ before periastron, on the near side after, and orbiting clockwise on the sky (see their figure~2).

Recently, \citet[][hereafter M12]{madura12} tightly constrained, for the first time, the 3-D orientation of \ec's orbit using a 3-D dynamical model for the broad, spatially-extended [\ion{Fe}{3}] emission observed by the \emph{HST} Space Telescope Imaging Spectrograph (STIS) \citep{gull09}. \citetalias{madura12} find that the observer's line-of-sight has an argument of periapsis $\omega \approx 240^{\circ}$ to $285^{\circ}$, with the binary orbital axis closely aligned in 3-D with the Homunculus polar axis at an inclination $i \approx 130^{\circ}$ to $145^{\circ}$ and position angle on the sky $\mathrm{PA}_{z} \approx 302^{\circ}$ to $327^{\circ}$, implying that \ecb\ indeed orbits clockwise on the sky.

Using a 3-D hydrodynamical model of \ec's binary colliding winds, we show in this letter that the orbital orientation derived by \citetalias{madura12} is consistent with the asymmetric NUV variability observed in the \emph{HST} ACS/HRC images of \ec\ presented in \citetalias{smith04b}. The model in this letter builds on the earlier work of \citetalias{smith04b}, but differs in a key respect, accounting for the wind-wind collision (WWC) cavity created by \ecb\ in \eca's dense wind (\citealt{pittard02, okazaki08, parkin09}; \citetalias{madura12}). This cavity reduces the H and \ion{Fe}{2} optical depths in line-of-sight to \eca\ (Groh 2011 and Groh et al. 2012, in preparation, hereafter G11 and G12, respectively), and causes a bore-hole effect \citep{maduraowocki10, madura10, madura11}, wherein the WWC cavity allows increased escape of continuum radiation from the hotter/deeper layers of \eca's extended wind photosphere at phases near periastron. The results of this letter provide insights into how/where NUV light escapes the \ec\ binary system and the time-dependent illumination of \ec's ejecta in various directions.

\section{Observations and Modeling} \label{obs}

We use the same difference images of \ec\ as shown in figure~1 of \citetalias{smith04b}, to which we refer the reader for further details. We examine the F220W filter NUV images \citep[probing the wavelength region from $\sim 1800 - 2600${\AA},][]{sirianni05} where the observed NUV excess is greatest. Each image frame shows the result of subtracting the average of all six observations from the original image at the indicated date. Here, we focus on the central $\pm 2''$ region where the observed asymmetric brightness changes are most pronounced \citepalias{smith04b}.

We zero the phase $\phi$ in \ec's 5.5-year cycle to JD = 2,452,819.8 with period $2022.7 \ \mathrm{days}$ \citep{damineli08a}. As discussed in \citetalias{smith04b}, most of the observed variability occurs $0.5''$ to $1''$ from the central stellar source, thus, $\phi$ must be corrected to take into account light travel time. The phase delay for regions $1''$ from \ec's central source, adopting $D = 2.3 \ \mathrm{kpc}$ \citep{smith06}, is $\Delta \phi \approx -0.007$. The revised phases corresponding to the individual difference images are therefore $\phi =$ 0.865, 0.925, 0.985, 0.003, 0.031, and 0.061. It is assumed for simplicity that phase zero of the spectroscopic cycle (the observations) coincides with phase zero of the orbital cycle (periastron). In a highly-eccentric binary system like \ec, the two values are not expected to be shifted by more than a few weeks (\citealt{groh10a}; \citetalias{madura12}), and so this assumption does not greatly affect our overall conclusions.

We use a 2-D radiative transfer code and a 3-D hydrodynamical model of \ec's colliding winds to interpret the phase-dependent \emph{HST} ACS/HRC F220W images. To determine the influence of \ecb, its low-density wind cavity, and the dense WWC-region walls on the observed spectrum of \ec, we use the 2-D radiative transfer models described in \citet{groh10a} and G11,G12. A 3-D Smoothed Particle Hydrodynamics (SPH) simulation is used to understand the effects of orbital motion on the WWC region formed between \eca\ and \ecb, and the spatial orientation of the WWC surface on the sky as a function of phase. The 3-D SPH simulation in this letter is identical to that used in \citetalias{madura12}, with the exception of the size of the computational domain, which here is a factor of ten smaller in order to focus on the dynamics of the inner WWC zone. The adopted stellar, wind, orbital, and orientation parameters are in Table~\ref{tab1}.

\begin{table}
\caption{Adopted Model Stellar, Wind, Orbital, and Orientation Parameters}
\label{tab1}
\begin{center}
\begin{tabular}{l c c}\hline
  Parameter & $\eta_{\mathrm{A}}$ & $\eta_{\mathrm{B}}$ \\ \hline
  Mass ($M_{\odot}$) & 90 & 30 \\
  Mass-Loss Rate ($M_{\odot}$ yr$^{-1}$) & $10^{-3}$ & $10^{-5}$ \\
  Wind Terminal Velocity (km s$^{-1}$) & 500 & 3000 \\
  Orbital Period (days) & \multicolumn{2}{c}{2024} \\
  Orbital Eccentricity $e$ & \multicolumn{2}{c}{0.9} \\
  Semi-major Axis Length $a$ (AU) & \multicolumn{2}{c}{15.4}  \\
  Orbital Inclination $i$ & \multicolumn{2}{c}{$\ 138^{\circ}$} \\
  Argument of Periapsis $\omega$ & \multicolumn{2}{c}{$\ 260^{\circ}$} \\
  Position Angle on Sky of Orbital Axis, $\mathrm{PA}_{z}$ & \multicolumn{2}{c}{$\ 312^{\circ}$} \\\hline
\end{tabular}
\end{center}
\end{table}

\begin{figure}
\centering
\resizebox{\hsize}{!}{\includegraphics{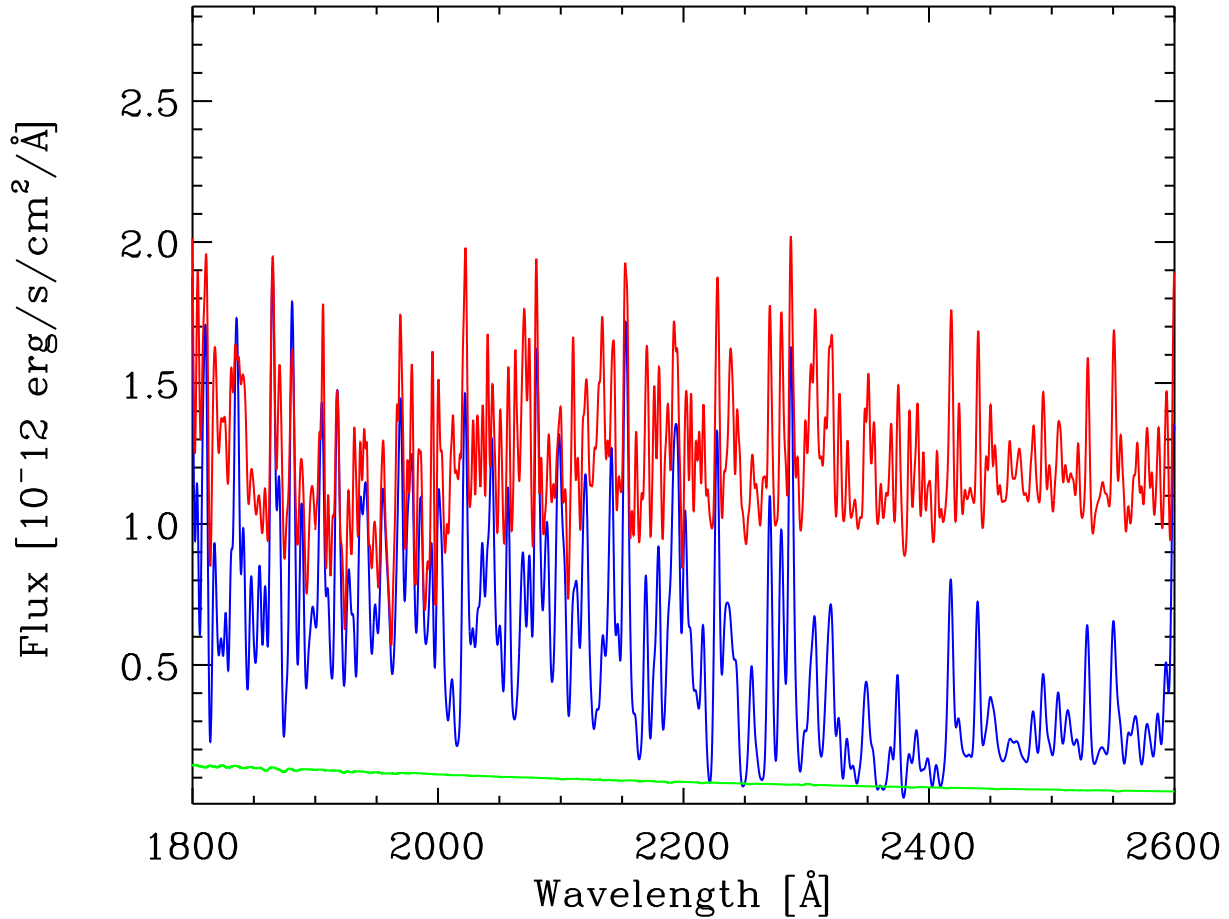}}
\caption{\label{fig1}{Comparison of model spectra in the range $1800 - 2600${\AA}, corresponding to the wavelength region probed by the \emph{HST} ACS/HRC F220W NUV images. All spectra have been smoothed to a resolution of $R = 600$ to make differences between models more clearly visible. \emph{Green Line}: Spectrum of \ecb\ computed using the 1-D CMFGEN model of \protect\citetalias{hillier06}. \emph{Blue Line}: Unmodified model spectrum of \eca\ from \protect\citetalias{hillier01, hillier06}. \emph{Red Line}: 2-D model spectrum from G12 at apastron, which includes the WWC cavity and its density-enhanced walls.}}
\end{figure}

\section{Origin of the near-UV Flux in the F220W Images} \label{ressub1}

Before examining the directional illumination of \ec's circumstellar ejecta, it is important to understand the physical origin of the observed NUV flux. According to the 2-D radiative transfer model of G11,G12, regions in line-of-sight to the wind cavity carved by \ecb\ are exposed to a larger NUV flux (red line of Figure~\ref{fig1}) than regions in line-of-sight to only the extended wind of \eca\ (blue line of Figure~\ref{fig1}). Between 1800 and 2600{\AA}, these modified wind models of \eca\ have a flux that is roughly an order of magnitude higher than the flux of \ecb. We assume for \ecb\ a temperature of 35,000 K and luminosity of $10^{6}$~\lsun\ (H06). Since $10^{6}$~\lsun\ is an upper limit for the luminosity of \ecb\ \citep{mehner10}, the flux contribution from \ecb\ could very likely be even less than what is shown in Figure~\ref{fig1}. Therefore, we find that \ecb\ does not contribute a substantial fraction of the observed F220W NUV flux in directions away from \eca\ and cannot explain the observed F220W NUV excess. Rather, we suggest that the NUV excess in the F220W images is regulated by the time-dependent nature of the WWC cavity carved by \ecb, which has two very important effects on \eca's extended wind photosphere.

First, lines-of-sight through the low-density WWC cavity have significantly reduced Fe~II optical depths (G11,G12). This reduction is especially pronounced in the F220W spectral region, which is full of Fe II lines (\citetalias{hillier06}; G12). Second, \eca's photosphere is extremely extended at UV wavelengths (\citetalias{hillier06}), leading to a significant bore-hole effect \citep{maduraowocki10, madura10, madura11} at phases around periastron, wherein the WWC cavity creates a hole, allowing increased escape of continuum radiation from the exposed hotter, deeper layers of \eca's photosphere.

Because \ecb\ is located well within the WWC cavity, $\lesssim$10\% of the F220W flux that reaches circumstellar ejecta exposed to the cavity comes from \ecb. Therefore, the time-dependent directional exposure of circumstellar material to excess F220W NUV flux most likely depends on the phase-dependent spatial orientation of the WWC cavity, and not just the obstruction of \ecb's NUV flux by \eca's wind.

\section{The Lighthouse Effect: Constraints on the Orbital Orientation Parameters} \label{ressub2}

Using our 3-D SPH model we investigated the phase-dependent spatial orientation of the WWC cavity for different orbital orientations with the goal of determining which orientations are capable of explaining the \emph{HST} F220W NUV images of \citetalias{smith04b}. For simplicity, each orientation discussed in this paper assumes that the orbital axis of the binary system is closely aligned in 3-D with the polar axis of the Homunculus nebula at an inclination $i = 138^{\circ}$ and position angle on the sky of $312^{\circ}$ \citepalias{madura12}.

A quantitative description of changes in the amplitude of the flux in the F220W images is deferred to future work as a 3-D radiative transfer code is necessary for such an analysis. Below, we compare the spatial location of features in the individual F220W images to those expected from the model based on the 3-D orientation of the WWC cavity opening. Compass directions (NE = north-east, etc.) are used in the descriptions below.

\begin{figure}
\centering
\resizebox{0.9\hsize}{!}{\includegraphics{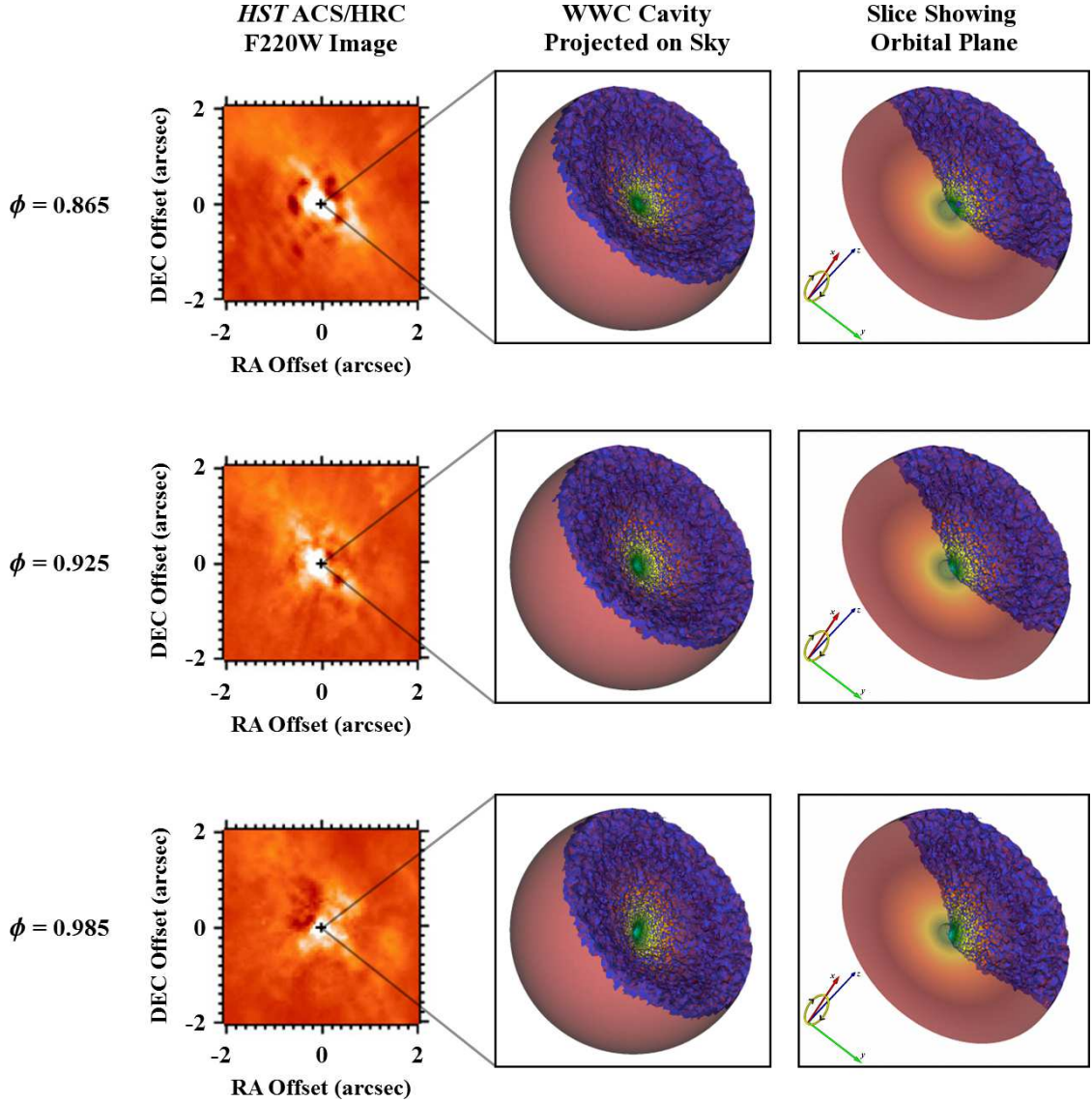}}
\caption{\label{fig2}{Comparison of observed \emph{HST} F220W images (left column, \citetalias{smith04b}) with the 3-D SPH model (right two columns) discussed in the text at phases (rows, top to bottom) $\phi = 0.865$, 0.925, and 0.985. Middle column: 3-D isovolume renderings of the SPH simulation assuming the parameters in Table~\protect\ref{tab1}, illustrating the orientation of the WWC cavity (dark purple) carved in \eca's extended wind (in red) as it would appear projected on the sky. The surface of the WWC cavity is color coded to radius, i.e. dark purple indicates material at larger radii ($\approx 75$ to $125$~AU) from the central stars, while green indicates material near the apex of the WWC region ($<15$~AU from \eca\ for these phases). Right column: Same as middle column, but with material below the orbital plane removed in order to show the complex dynamics of the inner WWC cavity. The inset in the lower left shows the orientation of the binary orbit (yellow) projected on the sky, along with the semi-major ($x$, red), semi-minor ($y$, green), and orbital ($z$, blue) axes. The black arrows indicate the clockwise orbital motion of the stars. North is up and east is left in all panels.}}
\end{figure}

\begin{figure}
\centering
\resizebox{0.9\hsize}{!}{\includegraphics{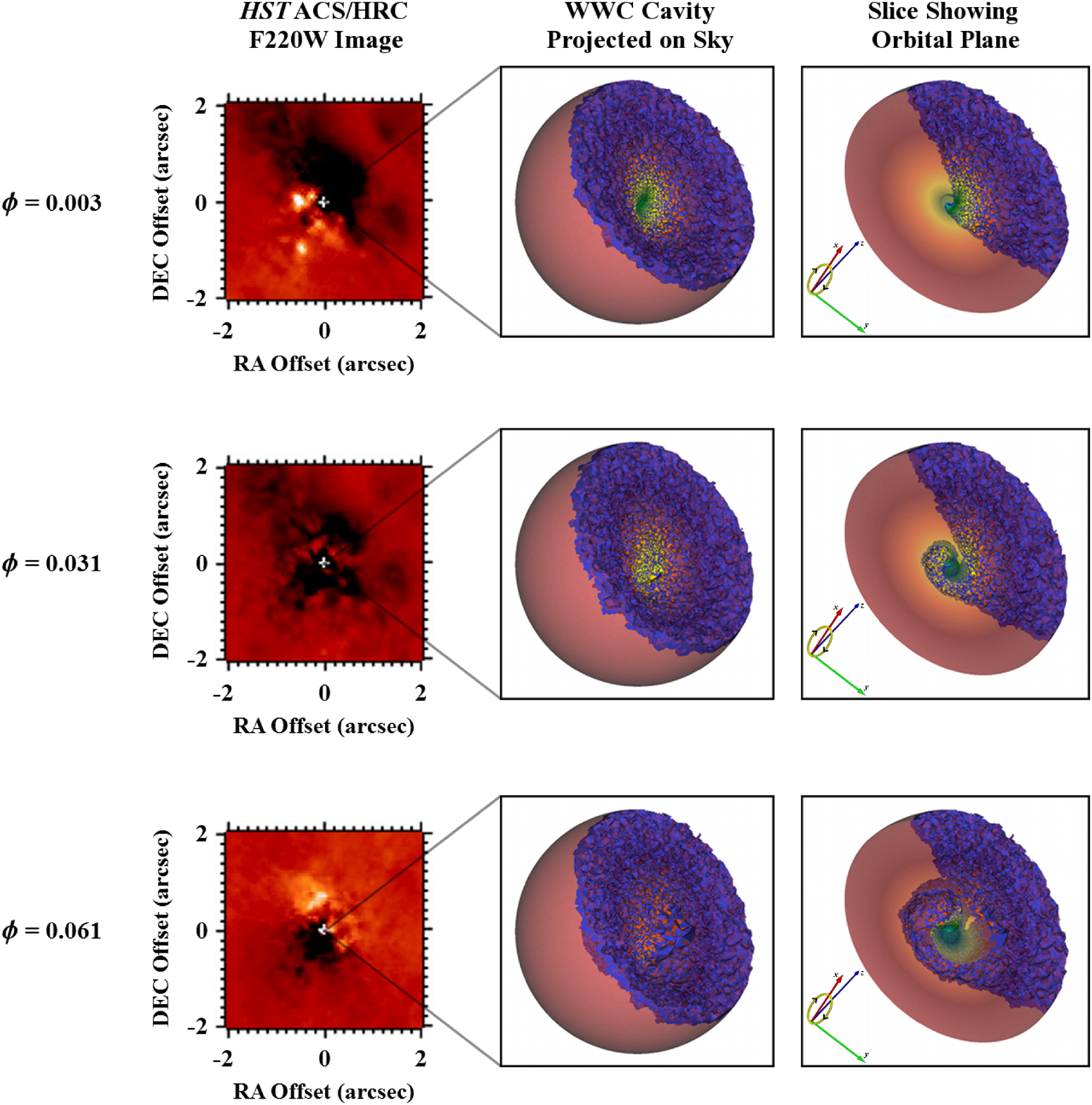}}
\caption{\label{fig3}{Same as Figure~\protect\ref{fig2}, but for phases (top to bottom) $\phi = 0.003$, 0.031, and 0.061.}}
\end{figure}

Figures \ref{fig2} and \ref{fig3} show that the binary orientation proposed by \citetalias{madura12} is capable of reasonably explaining the observed time-variable NUV `excess' emission seen in the F220W images. Hereafter, the terms `NUV excess' and `NUV deficit' refer to an increase and decrease, respectively, in the F220W NUV flux as compared to the average NUV flux of phases between 0.865 and 0.061.

At $\phi = 0.865$ and $0.925$, the observed NUV excess emission in the F220W images is brightest in regions in line-of-sight to the central stellar source and is elongated from NE to SW \citepalias{smith04b}. \ecb\ is far enough from \eca\ at these phases that orbital velocities are low and the WWC cavity maintains a simple axisymmetric cone-like shape (top two rows of Figure~2; also \citealt{okazaki08, parkin09, parkin11}; \citetalias{madura12}). For $\omega \approx 260^{\circ}$, the WWC cavity is open mostly toward the observer and is elongated in directions from the NE to the SW on the sky, causing the Fe~II optical depths in those directions to be reduced (G12). As a result, material in these directions should be exposed to a higher F220W NUV flux than the one-year average flux of phases between 0.865 and 0.061. The NUV excess thus occurs in these directions at these two phases because at later times the inner WWC cavity has a different 3-D spatial orientation (Figure~\ref{fig3}). Moreover, the models predict that because orbital velocities are low, the orientation of the WWC cavity should not change much between $\phi = 0.865$ and $0.925$, implying that the spatial orientation and amount of observed NUV excess should also not change much. The F220W images show that this is the case.

By $\phi = 0.985$ (bottom row of Figure~\ref{fig2}), the spatial orientation of the inner WWC cavity has changed, pointing more in directions to the SW on the sky. The WWC cavity at this phase is warped due to orbital motion. The F220W image shows an observed NUV excess in directions to the NW, SW, and SE, with darker, below-average-flux regions to the NE \citepalias{smith04b}. According to our model and interpretation, material to the NW, SW, and SE should have a NUV excess as it is exposed to a heavily modified primary wind. The darker region to the NE is likely due to there being more primary wind material in this direction at this phase compared to the average of all phases.

At $\phi = 0.003$, the inner WWC region has started to take on a spiral shape and points mainly in directions to the S on the sky and partly away from the observer (top row of Figure~\ref{fig3}). Interestingly, the F220W image shows NUV excesses to the S and E on the sky, with dark NUV-deficit regions to the N and W \citepalias{smith04b}. At $\phi = 0.003$, there is a significant bore-hole effect concentrated in directions to the S on the sky. The primary wind is carved to the S as well, which should allow the increased flux from the bore-hole to reach material in this direction. Since there is no bore-hole effect to the S at the other phases, we expect a NUV excess in material there at $\phi = 0.003$. Similar reasoning explains why there is also a NUV deficit to the N and W.

At $\phi = 0.031$ (middle row of Figure~\ref{fig3}), the inner $\pm 2''$ region of the F220W image is very dark \citepalias{smith04b}. The inner WWC cavity gets highly distorted and the additional NUV radiation that would escape from the inner layers of \eca\ becomes embedded and trapped by its dense wind \citepalias{madura12}. Therefore, a NUV deficit and a large dark region appears in the F220W difference image.

By $\phi = 0.061$ (bottom row of Figure~\ref{fig3}), orbital speeds have decreased and the WWC cavity has started to increase in size in directions to the NNE on the sky, significantly carving the wind of \eca\ in this direction. Consequently, one would expect a significant decrease in the amount of Fe~II absorption in directions to the NNE compared to the average, and thus an observed NUV excess. There is also more wind material from \eca\ in directions to the S compared to the average, which should cause a NUV deficit. The F220W image shows that there is indeed an observed NUV excess to the NNE and a dark region to the S \citepalias{smith04b}.

\begin{figure}
\centering
\resizebox{\hsize}{!}{\includegraphics{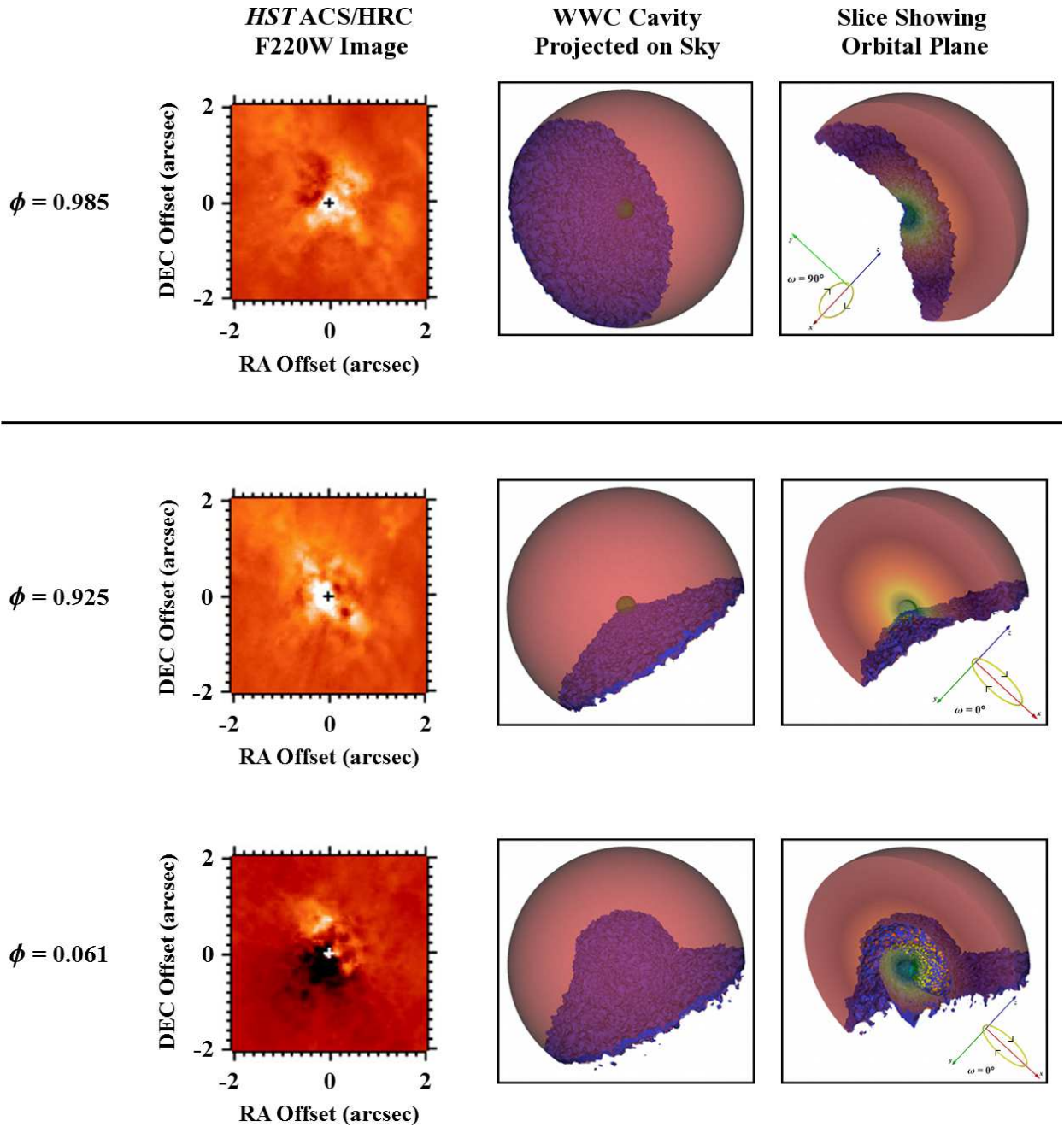}}
\caption{\label{fig4}{Similar to Figures~\protect\ref{fig2} and \protect\ref{fig3}, but for orbital orientations assuming $\omega = 90^{\circ}$ (top row) and $\omega = 0^{\circ}$ (bottom two rows) at selected phases. The red wind of \eca\ in the renderings of the last two columns has been made semi-transparent in order to allow the reader to see the dark purple surface of the WWC region, which opens away from the observer for these two orientations.}}
\end{figure}

We find that only binary orientations with $230^{\circ} \lesssim \omega \lesssim 15^{\circ}$ are capable of reasonably explaining the \emph{extended} ($\sim 0.2''$ to $2''$) NUV excess emission observed in the F220W images. Other orientations appear to have extreme difficulty explaining the NUV variability. For example, at an orientation of $\omega = 90^{\circ}$ (top row of Figure~\ref{fig4}), during most of the binary orbit, the WWC cavity is open and pointing away from the observer and to the E/SE. Thus, between $\phi = 0.865$ and $0.985$, the observer should detect the NUV radiation field from the unmodified wind of \eca\ on the central stellar source, and a possible NUV excess to the E/SE. At $\phi = 0.061$, due to the wrapping of the WWC region, a NUV excess would be expected to the S. Yet, the spatial location of the observed bright and dark spots is nearly the exact opposite.

However, we find that the NUV variability on the central stellar source (central $\pm 0.2''$) additionally constrains the orbital orientation to values of $\omega \approx 230^{\circ}$ to $315^{\circ}$. The F220W images at $\phi = 0.865$, 0.925, and 0.985 show an observed NUV excess on the central stellar source, while $\phi = 0.003$, 0.031, and 0.061 show a deficit (\citetalias{smith04b}; see also \citealt{martin06}; \citealt{mehner11}). In order to have a NUV excess on the central source before periastron, the WWC cavity should be open mostly toward the observer at these phases. NUV deficits after periastron are due to increased amounts of primary wind material (compared to the average) at these times. This appears to be the case for $\omega \approx 230^{\circ}$ to $315^{\circ}$. In contrast, assuming $\omega = 0^{\circ}$, for example, the WWC cavity opens away from the observer at phases $\phi = 0.865$ to 0.003 (Figure~\ref{fig4}). At $\phi = 0.061$, the WWC spirals in between the observer and \eca (bottom row Figure~\ref{fig4}). This would lead one to expect a NUV deficit before periastron and an excess after if $\omega = 0^{\circ}$, which is not seen in the observations.

Moreover, binary orientations that lie significantly outside the range $\omega \approx 230^{\circ}$ to $315^{\circ}$ have great difficulty explaining other multi-wavelength diagnostics of \ec\ since such orientations place the observer's line-of-sight through \eca's optically-thick wind for most of the orbital period (\citealt{pittard02, okazaki08, parkin09, parkin11, groh10b}; \citetalias{madura12}; G11,G12). In contrast, the 3-D orientation and direction of orbit proposed by \citetalias{madura12} appears consistent with all known observations of \ec\ to date, including those analyzed here.

The results of this letter go well beyond constraining the orientation of \ec's binary orbit. It is clear that the motion of the WWC cavity as \ecb\ moves about in its highly elliptical orbit leads to an important `lighthouse effect' in \ec, wherein circum-binary ejecta is exposed to a time- \emph{and} direction-dependent modified NUV radiation field of \eca\ caused by a bore-hole effect and decrease in Fe II optical depths. This lighthouse effect is crucial for understanding how NUV light escapes the \ec\ system and the phase-dependent illumination of distant ejecta in different directions.

The lighthouse effect also provides a valuable diagnostic for helping constrain the exact timing of periastron. Based on the available observations of \citetalias{smith04b} and the simple model in this letter, periastron should occur between $\phi = 0.985$ and $\phi = 0.031$, most likely very close to $\phi = 0.003$ since this is when some NUV excess is still visible to the SE in the difference images (i.e. before distortion of the WWC cavity causes the NUV radiation to be trapped by \eca's dense wind).

Future spatially-resolved observations with better time sampling, together with detailed 3-D radiative transfer models, can help place much tighter constraints on the exact timing of periastron, and possibly the orbital eccentricity. We emphasize that future observations of \ec\ should focus not only on phases around $\phi = 0$, but also on the extended recovery period up until $\phi \approx 0.2$, during which time the WWC cavity is increasing in size and reestablishing its axisymmetric cone-like shape. Future monitoring of these phases is crucial for determining when various forms of radiation can escape via the WWC cavity in directions away from \eca.

\section*{Acknowledgements}
We wish to thank S. Owocki, N. Smith, and T. Gull for interesting discussions relevant to this work. We thank D.~J. Hillier for making available, and providing continuous support for, the CMFGEN and 2-D \citet{busche05} radiative transfer codes, and the models of \eca\ and \ecb. We also thank C. Kruip for assisting in the 3-D visualization of the SPH code output, and the MPG for financial support.

\end{document}